\documentclass[aps,amsfonts,superscriptaddress, nofootinbib, floatfix, 10pt,twocolumn,prd]{revtex4-1} 
\usepackage{graphicx}
\usepackage{amsmath}
\usepackage[latin1]{inputenc}
\usepackage{amsbsy}
\usepackage{color}
\usepackage{epsfig}
\usepackage{graphicx}
\usepackage{amsmath}
\usepackage{caption}
\usepackage{subcaption}
\usepackage{amssymb}
\usepackage{bbm}
\usepackage{amsbsy}
\usepackage{slashed}

\usepackage{xcolor}

\usepackage{soul}

\usepackage[normalem]{ulem}  

\def\be{\begin{equation}}

\def\ee{\end{equation}}
\def\bea{\begin{eqnarray}}
\def\eea{\end{eqnarray}}

\begin{document} 


\title{Thermo-magnetic corrections to  $\pi$-$\pi$ Scattering Lengths in the Linear Sigma Model}

\author{M.~Loewe}
\email{mloewe@fis.puc.cl}
\affiliation{Instituto de F\'isica, Pontificia Universidad Cat\'olica de Chile, Casilla 306, San\-tia\-go 22, Chile}
\affiliation{Centre for Theoretical and Mathematical Physics and Department of Physics,
University of Cape Town, Rondebosch 7700, South Africa}
\affiliation{Centro Cient\'ifico Tecnol\'ogico de Valpara\'iso-CCTVAL, Universidad T\'ecnica Federico Santa Mar\'ia, Casilla 110-V, Valpara\'iso, Chile} 
\author{E. Mu\~noz}
\email{munozt@fis.puc.cl}
\affiliation{Instituto de F\'isica, Pontificia Universidad Cat\'olica de Chile, Casilla 306, San\-tia\-go 22, Chile}
\author{R. Zamora}
\email{rzamorajofre@gmail.com}
\affiliation{Instituto de Ciencias B\'asicas, Universidad Diego Portales, Casilla 298-V, Santiago, Chile} 
\affiliation{Centro de Investigaci\'on y Desarrollo de Ciencias Aeroespaciales (CIDCA), Fuerza A\'erea de Chile, Santiago 8020744, Chile}



\begin{abstract}
In this article, we extend our previous study of the $\pi$-$\pi$ scattering lengths under the presence of an external
magnetic field, including finite temperature effects. The novelty of this work is precisely the introduction of
temperature into the discussion, and its interplay with the magnetic field. As in the previous article, we base our
analysis in the linear sigma-model, and our calculations are exact within this context. Although the effects
are comparatively small, it is interesting to remark that magnetic field and temperature display opposite
effects over the scattering lengths.
\end{abstract}


\maketitle


\section{Introduction}

An interesting and relevant question in the field of heavy ion collision experiments is to search
for the possibility of disentangling temperature and magnetic effects. In this article we will refer
to $\pi-\pi$ scattering lengths, extending our previous results  \cite{PRDT0}  at zero temperature
and finite magnetic fields. In this work, we shall present analytical results that fully capture
both thermal and magnetic effects without any restriction or special ordering in these
parameters. For this purpose, our calculations will be based on the linear sigma model
using techniques associated to the spectral density functions that allows us to handle
the infinite series over Landau levels in a closed analytical form. Infinite Matsubara sums
were also performed exactly with complex contour integration techniques.

The paper is organized as follows: In section II, we introduce the Linear sigma model and the general
formalism for scattering lengths. In section III, the detailed Feynman diagrams are displayed, using
the Schwinger propagators for finite magnetic fields. The two master integrals required for the
calculation of all the diagrams are formulated. Therein, we also display the analytical results
for the scattering lengths, with mathematical details deferred to Appendixes. Finally, in section IV
we show our numerical results, discussing the combined effects of temperature and magnetic field. 
 
\section{Linear sigma model and $\pi$-$\pi$ scattering} \label{secII}

In our previous article \cite{PRDT0} we studied the  $\pi$-$\pi$ scattering lengths within the context of the linear sigma model
in the phase where the chiral symmetry is
broken, 
\begin{eqnarray}
\mathcal{L}&=&\bar{\psi}\left[i\gamma^{\mu}\partial_{\mu}-m_{\psi}-g(\sigma+i\vec{\pi}\cdot\vec{\tau}\gamma_{5})\right]\psi\\
&&+\frac{1}{2}\left[(\partial\vec{\pi})^2+m_{\pi}^2\vec{\pi}^2\right]+\frac{1}{2}\left[(\partial\sigma)^2+m_{\sigma}^2 \sigma^2\right]\nonumber\\
&&-\lambda^2v\sigma(\sigma^2+\vec{\pi}^2)-\frac{\lambda^2}{4}(\sigma^2+\vec{\pi}^2)^2+(\varepsilon c-vm_{\pi}^2)\sigma.\nonumber
\end{eqnarray}

The model includes a doublet of Fermi fields, which in our case will be ignored since they represent nucleons which are too heavy
as compared with the scalar sigma meson and the relevant energy scale. The role of the sigma meson is to break explicitly the
chiral symmetry. Then, as usual, the sigma field is expanded around its expectation value and its not difficult to see that 
$m_{\psi}=gv$, $m_{\pi}^2=\mu^2+\lambda^2v^2$ and
$m_{\sigma}^2=\mu^2+3\lambda^2v^2$. For details see \cite{Lee}.
 Perturbation theory at the
tree level allows us to identify the pion decay constants as
$f_{\pi}=v$. Finite temperature effects on this model have been studied by several authors, discussing the thermal
evolution of masses, $f_\pi(T)$, the effective potential, etc.~\cite{cep,renor,masa,super,magn,inverse,sigma,Loewe,Larsen,Bilic,Petropolus,wagner,kovacs1,kovacs2,kovacs3}.

The most general decomposition for the scattering amplitude for particles with definite isospin quantum numbers is~\cite{Collins, Gasser}

\begin{eqnarray}
T_{\alpha\beta;\delta\gamma}&=&A(s,t,u)\delta_{\alpha\beta}\delta_{\delta\gamma}+A(t,s,u)\delta_{\alpha\gamma}\delta_{\beta\delta}\nonumber\\
&&+A(u,t,s)\delta_{\alpha\delta}\delta_{\beta\gamma},
\label{proyectores}
\end{eqnarray}
\noindent where $\alpha$, $\beta$, $\gamma$, $\delta$ denote
isospin components.

By using appropriate projection operators, it is possible to find
the following isospin dependent scattering amplitudes



\begin{align}
T^{0}&=3A(s,t,u)+A(t,s,u)+A(u,t,s),\label{eq3}\\
T^{1}&=A(t,s,u)-A(u,t,s),\label{eq4}\\
T^{2}&=A(t,s,u)+A(u,t,s),
\label{eq5}
\end{align}

\noindent where $T^I$ denotes a scattering amplitude in a given isospin channel $I = \{0,1,2\}$.\\

As it is well known~\cite{Collins}, below the inelastic threshold any scattering amplitude can be expanded
in terms of partial amplitudes which can be parametrized by the phase shifts for each angular 
momentum channel $\ell$. Therefore, in the low-energy region the isospin dependent
scattering amplitude can be expanded in partial wave components $T_\ell^I$. The real part of this amplitude
\begin{equation}
\Re\left(T_{\ell}^{I}\right)=\left(\frac{p^{2}}{m_{\pi}^{2}}\right)^{\ell}\left(a_{\ell}^{I}+\frac{p^2}{m_{\pi}^{2}}b_{\ell}^{I}+\ldots\right)
\end{equation}
is expressed in terms of the
scattering lengths $a_{\ell}^{I}$, and the scattering slopes $b_{\ell}^{I}$, respectively.
The scattering lengths satisfy the hierarchy $|a_{0}^{I}|>|a_{1}^{I}|>|a_{2}^{I}|...$.
In particular, in order to obtain the scattering lengths $a_0^I$, it is
sufficient to calculate the scattering amplitude $T^I$ in the static
limit, i.e. when $s \to 4m_\pi^2$, $t\to 0$ and $u\to 0$
\begin{equation}
a_{0}^{I}=\frac{1}{32\pi}T^{I}\left(s \to 4m_{\pi}^2,t\to 0, u\to0\right).\label{eq:a0I}
\end{equation} 
 The first measurement of $\pi $-$\pi $ scattering lengths was carried on by Rosellet 
{\em et al.}
~\cite{Rosellet}. More recently, these parameters have been measured  using pionium atoms in the DIRAC experiment \cite{DIRAC} and also through the decay of heavy quarkonium states into $\pi $-$\pi $ final states where the so called cusp-effect was found~\cite{quarkonium}. In
our recent work \cite{PRDT0}, we obtained an exact analytical expression for Eq.~(\ref{eq:a0I}) for $I=0,2$ in a background magnetic field of arbitrary strength, at zero temperature. In this article,
we extend our previous result to include finite temperatures. Therefore, in what follows we shall present the exact expression for the scattering lengths Eq.(\ref{eq:a0I}) at arbitrary magnitudes of the external magnetic field and temperature.

\section{Scattering lengths at finite magnetic field and finite temperature}
In a previous work, based on a perturbative treatment of the bosonic Schwinger propagator valid for small magnetic fields, some of us discussed the magnetic dependence of the $\pi$-$\pi$ scattering lengths within the context of the linear sigma model~\cite{Leandro}.
We found that this magnetic evolution displays an opposite trend with respect to thermal corrections on the scattering lengths, in agreement with the literature \cite{TLSM}. At low magnetic field intensities, the scattering lengths in the isospin channel $I = 2$ increase, whereas their projection into the channel $I = 0$ diminishes, both as functions of the magnetic field.
More recently, we extended the analysis in the zero temperature scenario, thus obtaining exact analytical expressions valid in the whole range of magnetic field intensities \cite{PRDT0}.
\bigskip

\begin{figure}[h!]
{\centering
\includegraphics[scale=0.75]{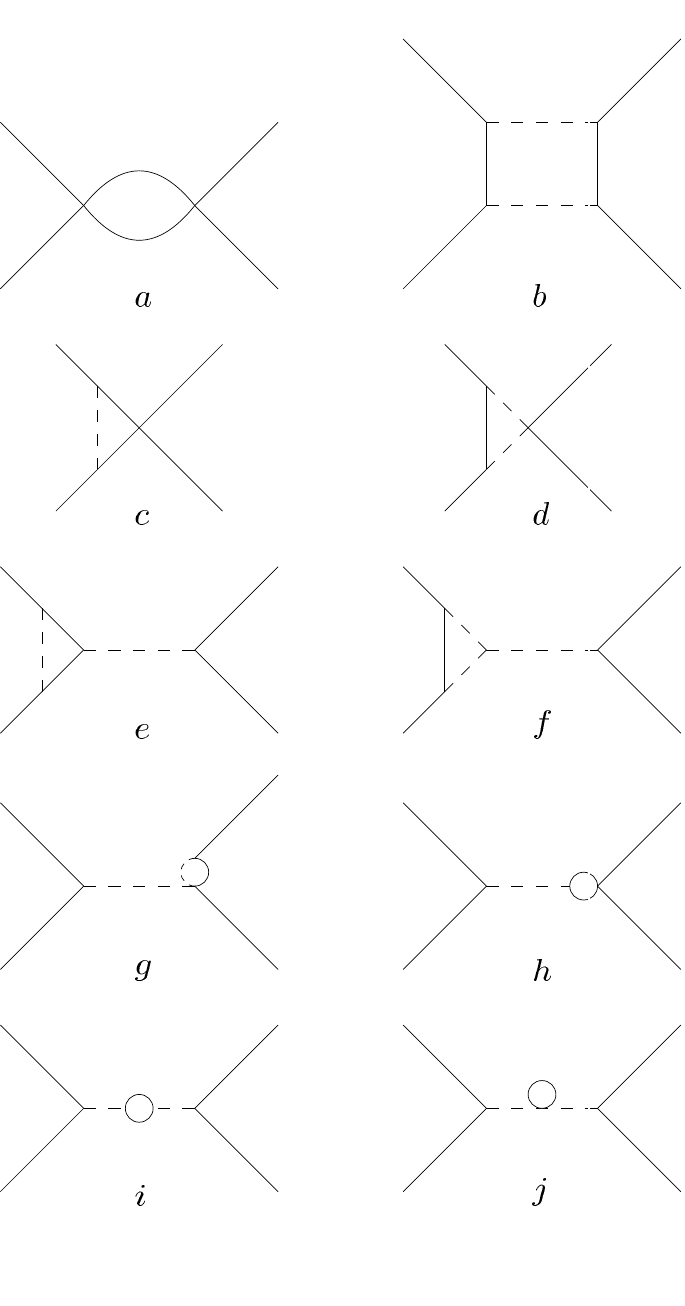}}
\caption{One-loop diagrams relevant to the $\pi$-$\pi$ scattering lengths}. Continuous and dashed lines represent pions and $\sigma$ mesons respectively.
\label{oneloop}
\end{figure}
\noindent In the linear sigma model, the relevant diagrams that contribute to $\pi$-$\pi$ scattering are shown in Fig. \ref{oneloop}. Notice that tadpole-like diagrams associated to mass corrections of the sigma field, do not contribute to the  $\pi$-$\pi$ scattering amplitudes, since their imaginary part vanishes. These tadpoles are extremely small in the limit of a very large mass of the sigma field. This approximation is valid since, as we know, $m_{\sigma} \approx 550$~MeV is much larger than the pion mass. Fermions, i.e. nucleons that may interact with pions, are not considered in our discussion. As a consequence, the sigma field propagator is contracted to a point.

\bigskip
\begin{figure}
        \begin{subfigure}[b]{0.5\textwidth}
		\centering
                \includegraphics[scale=0.6]{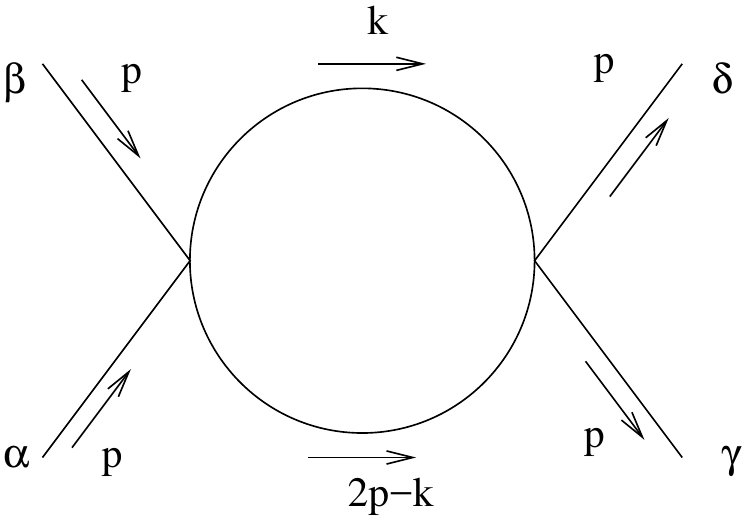}
                \caption{$s$-channel diagram.}
                \label{fig:gull}
        \end{subfigure}
        \begin{subfigure}[b]{0.5\textwidth}
		\centering
                \includegraphics[scale=0.6]{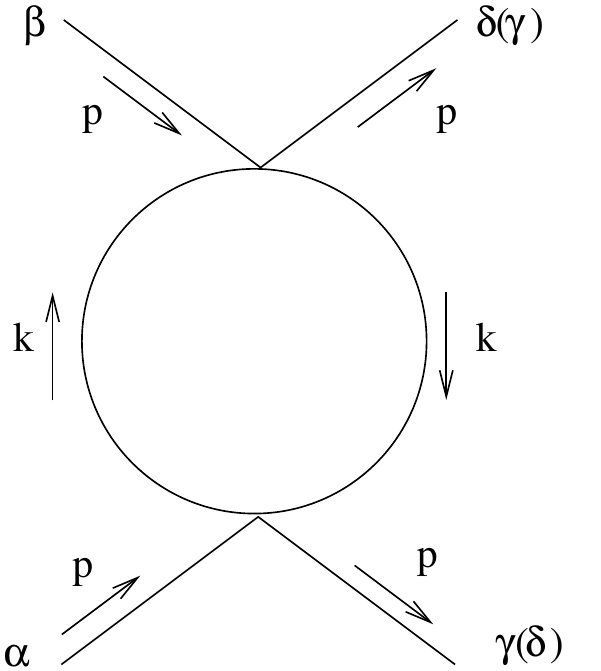}
                \caption{$t$ and $u$ channel diagram.}
                \label{fig:gull2}
        \end{subfigure}
        \caption{``Fish-type'' diagrams.}\label{fig:a diagrams}
\end{figure}

\noindent From these considerations, we see that all relevant diagrams reduce to a horizontal ($s$-channel) or vertical ($t$ and $u$ channels) {\em``fish-type"} pion loops contributions, as shown in Fig.~\ref{fig:a diagrams}. Then, we need to compute such diagrams as a function of the magnetic field intensity and finite temperature. This is an interesting problem, not only because of physical implications, but also due to new analytical results that we shall present below.

\bigskip
\noindent Let us derive our starting expression for the bosonic propagator as a sum of Landau levels \cite{Ayalaetal,PRDT0}. The bosonic Schwinger propagator for a charged pion of charge $q$ 
subject to a uniform magnetic field along the third spatial coordinate, in the proper time representation is given by
\begin{align}
iD^{B} (k)&=\int_0^\infty \frac{ds}{\cos(qBs)}e^{ is\left(k_{||}^2-k_\perp^2\frac{\tan(qBs)}{qBs}-m_\pi^2 +i\epsilon \right)}.
\end{align}
 
\noindent After inserting this propagator in the  fish-type diagrams, one finds that all contributions reduce to two types of integrals \cite{PRDT0}
\begin{eqnarray}
I_1[B,p_0]	&=&\int\frac{d^4k}{(2\pi)^4}iD^{B}(k_0,\mathbf k)iD^{B}(k_0-2p_0,\mathbf k), \label{int1}\\
I_2[B]	&=& \int\frac{d^4k}{(2\pi)^4} \left[iD^{B}(k_0,\mathbf k)\right]^2 \label{int2}.
\label{eq_I12}
\end{eqnarray}

For technical purposes, we shall calculate the integrals with the expression for the propagator at finite magnetic field in terms of Landau levels, as presented in \cite{Ayalaetal}.
\begin{eqnarray}
i D^{B}(k) = 2  \sum_{l=0}^{\infty}(-1)^{l} L_{l}\left(\frac{2 k_{\perp}^2}{qB} \right) e^{-k_{\perp}^2/qB} i\Delta_{l}^{B}(k_{\parallel}),
\label{eq_prop}
\end{eqnarray}
where $L_{l}(z)$ are the Laguerre polynomials, and 
we have defined the effective ``parallel'' propagators
\begin{eqnarray}
i\Delta_{l}^{B}(k_{\parallel}) = \frac{i}{k_{\parallel}^{2} - (2l + 1)qB - m_{\pi}^2 + i\epsilon }.
\label{eq_prop_Land}
\end{eqnarray}

The above expressions correspond to the zero-temperature scenario. However, it is straightforward to generalize them
to the finite temperature case by analytic continuation into Matsubara frequency space, i.e.
\begin{eqnarray}
k_0 \rightarrow i\omega_{n} = 2\pi n/\beta,&n\in Z,\nonumber\\
p_0 \rightarrow i\nu_{m} = 2\pi m/\beta, & m\in Z
\end{eqnarray}
where $\beta=1/T$, and the corresponding substitution of the integral in $k_0$ by a sum,
\begin{eqnarray}
\int \frac{d^2 k_{\parallel}}{(2\pi)^2} \rightarrow \frac{i}{\beta} \sum_{n\in Z} \int_{-\infty}^{+\infty} \frac{dk_3}{2\pi}
\end{eqnarray}

Let us first consider the calculation of $I _{1}[T,B,p_0]$, after its definition in Eq.(\ref{eq_I12}), substituting the infinite series for the propagators, Eq.(\ref{eq_prop}), we are lead to
\begin{eqnarray}
&&I_1[T,B,i\nu_{m}] = \frac{i}{\beta}\sum_{n\in Z}\int \frac{d^3 k}{(2\pi)^3} i D^{B}(i\omega_n,\mathbf{k}) \nonumber \\
&\times& iD^{B}(i\omega_n - 2i\nu_m,\mathbf{k})\nonumber \\
&& = 4 \sum_{l=0}^{\infty} \sum_{l' = 0}^{\infty}(-1)^{l + l'} 
G_{l,l'}(i\nu_m) \nonumber \\
&&\times \left[\int\frac{d^2 k_{\perp}}{(2\pi)^2} e^{-2\,k_{\perp}^2/qB}  L_{l}\left(\frac{2 k_{\perp}^2}{qB} \right) L_{l'}\left(\frac{2 k_{\perp}^2}{qB} \right)\right].\nonumber\\
\label{eq_double_sum}
\end{eqnarray}
Here, we have defined the functions
\begin{eqnarray}
G_{l,l'}(T,i\nu_m) &=& \int_{-\infty}^{\infty}\frac{dk_3}{2\pi} \frac{i}{\beta}\sum_{n\in Z} i\Delta_{l}^{B}(i\omega_n,k_3)\nonumber\\
&&\times i\Delta_{l'}^{B}(i\omega_n - 2i\nu_{m},k_3). 
\end{eqnarray}

Let us now calculate the integral over the Laguerre polynomials in the second term, by using 2-dimensional ``spherical coordinates",
with $0 \le | k_{\perp}| < \infty$,
\begin{eqnarray}
d^{2}k_{\perp} &=& 2	\pi |k_{\perp}| d| k_{\perp}| = \frac{\pi\,q\,B}{2} dx,
\end{eqnarray}
where we have defined the auxiliary variable $x = 2 k_{\perp}^2/qB$, with $0 \le x < \infty$. Therefore, we have
\begin{eqnarray}
&&\int\frac{d^2 k_{\perp}}{(2\pi)^2} e^{-2\,k_{\perp}^2/qB}  L_{l}\left(\frac{2 k_{\perp}^2}{qB} \right) L_{l'}\left(\frac{2 k_{\perp}^2}{qB} \right) \nonumber \\
&=& \frac{1}{4\pi^2}\frac{\pi\,q B}{2} \int_{0}^{\infty} dx e^{-x} L_{l}(x) L_{l'}(x)\nonumber\\
&=& \frac{qB}{8\pi}\delta_{l,l'},
\label{eq_Laguerre_orto}
\end{eqnarray}
where the orthogonality relation between Laguerre polynomials was used.
Substituting this result into Eq.(\ref{eq_double_sum}), we end up with the expression
\begin{eqnarray}
I_1[T,B,i\nu_m] = \frac{qB}{2\pi}\sum_{l=0}^{\infty} G_{l,l}(T,i\nu_m).
\label{eq_I1_reduced}
\end{eqnarray}

As shown in detail in Appendix, we calculate $G_{l,l}(i\nu_m)$ by first integrating over $k_0$ in the complex plane, and later over $k_3$. This procedure allows us to obtain the infinite series
\begin{eqnarray}
I_1[T,B,i\nu_m] &=& \frac{qB}{4\pi (i\nu_m)} \nonumber \\
&\times& \Im m\,\sum_{l=0}^{\infty}  \int_{-\infty}^{+\infty}\frac{dk_3}{2\pi}g(T,E_l(k_3),i\nu_m), \nonumber \\
\label{eq_I1_reduced}
\end{eqnarray}
with $E_l(k_3) = \sqrt{k_3^2 + m_{\pi}^2 + q B(2 l + 1)}$, and
the functions $g(T,E,i\nu_m)$ defined by 
\begin{eqnarray}
g(T,E,i\nu_m) = \frac{\coth\left(\beta E/2 \right)}{E[E - i\nu_m]}.
\end{eqnarray}

As discussed in Appendix, we introduce the spectral density
\begin{eqnarray}
\rho(E) = \sum_{l=0}^{\infty}\int_{-\infty}^{\infty}\frac{dk_3}{2\pi}\delta\left( E - E_l(k_3)  \right),
\end{eqnarray}
such that Eq.(\ref{eq_I1_reduced}) can be expressed as single integral over the energy domain
\begin{eqnarray}
I_1[T,B,i\nu_m] = \frac{qB}{4\pi (i\nu_m)} \Im m\, \int_{0}^{\infty} dE \rho(E) g(T,E,i\nu_m), \nonumber \\
\end{eqnarray}
where a closed analytical expression for the spectral density was derived in Appendix,
\begin{eqnarray}
\rho(E) &=& \frac{E}{\pi\sqrt{2 q B}}\Theta\left(E - \sqrt{m_{\pi}^{2} + qB } \right)\nonumber\\
&&\times \left\{
\zeta\left(\frac{1}{2},\frac{E^2 - m_{\pi}^2 - qB}{2 q B}-\Bigg\lfloor \frac{E^2 - m_{\pi}^2 - qB}{2 q B} \Bigg\rfloor \right)\right.\nonumber\\ 
&&\left.- \zeta\left(\frac{1}{2},\frac{E^2 - m_{\pi}^2 - qB}{2 q B} + 1 \right)
\right\},
\end{eqnarray}
with $\lfloor z \rfloor$ the integer part of $z$ and $\zeta(s,z)$ the Hurwitz Zeta function.
\noindent In order to obtain the scattering lengths $a_0^I$, we use the decomposition of the scattering amplitude in the different isospin channels presented in Section \ref{secII}. 
Since we are only interested in the scattering lengths $a_0^I$, it is enough to calculate the scattering amplitude in the static limit. Therefore, we normalize by the experimental values at tree level ($a_0^0(\text{exp})=0.217$ and $a_0^2(\text{exp})=-0.041$),  to obtain the expressions
\begin{eqnarray}
a_0^0(T,B)	&=&a_0^0(\text{exp})+\frac1{32\pi}\bigl(3A(s,t,u)+A(t,s,u)\nonumber \\
&+& A(u,t,s)\bigr),\nonumber\\
a_0^2(T,B)	&=&a_0^2(\text{exp})+\frac1{32\pi}\left(A(t,s,u)+A(u,t,s)\right).
\end{eqnarray}

Here, $A(s,t,u)$, $A(t,s,u)$ and $A(u,t,s)$ correspond to all $s$-channel, $t$-channel and $u$-channel contributions, respectively.
On the other hand, the $s$-channel contribution is obtained from $I_1[T,B,i\nu_m \rightarrow p_0=m_{\pi}]$, while those for the $t$- and $u$-channels are obtained from $I_2[T,B]=I_1[T,B,i\nu_m \rightarrow p_0=0]$, according to the following expressions

\begin{eqnarray}
A(s,t,u) &=& -4\lambda^4\left(1 - \frac{12 \lambda^2 v^2 }{m_{\sigma}^2}+\frac{24 \lambda ^4 v^4 }{m_{\sigma }^4}\right) \nonumber \\
&\times& I_1[T,B,m_{\pi}],\nonumber\\
A(t,s,u) + A(u,t,s) &=& -8 \lambda^4\left(
1 - \frac{12 \lambda^2 v^2}{m_{\sigma}^2} + \frac{24 \lambda ^4 v^4 }{m_{\sigma }^4}
\right) \nonumber \\
&&\times I_1[T,B,0].
\end{eqnarray}

The experimental values in the absence of magnetic field $B = 0$ are given by \cite{peyaud}  $a_0^0(\text{exp})=0.217$ and $a_0^2(\text{exp})=-0.041$. The mass for the sigma meson
is set to  $m_{\sigma} = 550$ MeV, and the mass for the pion $m_{\pi} = 140$ MeV, with the parameter $v = 89$ and $\lambda^2=4.26$.

\section{Results and Conclusions}

We have presented a novel method to calculate the scattering lengths for $\pi-\pi$ scattering within the linear sigma model at the one-loop level, in the isospin channels $I = \{0,2\}$, as functions of the temperature and external magnetic field intensity. 
We show three plots of the calculation of the scattering lengths $a_0^0$ and $a_0^2$ as a temperature for three magnetic field values. From both figures, it is clear that at finite magnetic field, while $a_0^0$ increases as a function of temperature, $a_0^2$ displays the opposite trend.
It is interesting to remark that already the presence of a small magnetic field is sufficient to invert the temperature dependence at zero magnetic field \cite{TLSM,Ruiz}. This effect can be explained by the symmetry of the scattering length parameters, where
$a_0^0$ arises from the trace of the tensor, which is a scalar, whereas the isospin 2 channel corresponds to the most symmetric realization. Qualitatively, the magnetic moment of the semiclassical orbits associated to the Landau levels tend to align along the magnetic field direction, while
temperature tends to randomize those directions. Therefore, under finite magnetic fields the most symmetric state enhances its scattering rate.

\begin{figure}[h!]
{\centering
\includegraphics[scale=0.25]{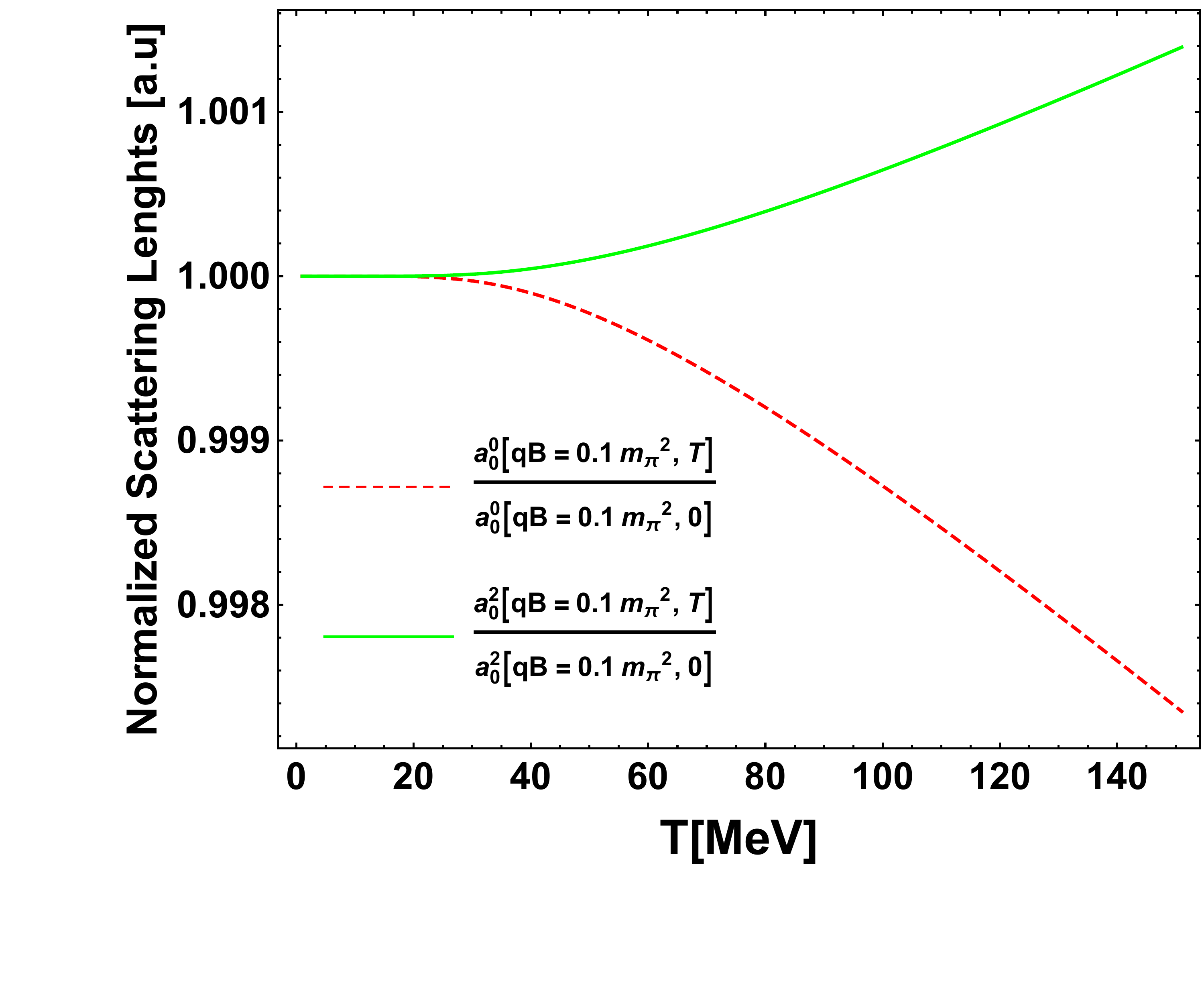}}
\caption{(Color online) The scattering parameters $a_0^{0}(B,T)/a_0^{0}(B,0)$ (dashed) and $a_0^{2}(B,T)/a_0^{2}(B,0)$ (solid) are displayed as a function of the temperature for $qB=0.1 m_{\pi}^2$.}
\label{plot1}
\end{figure}

\begin{figure}[h!]
{\centering
\includegraphics[scale=0.25]{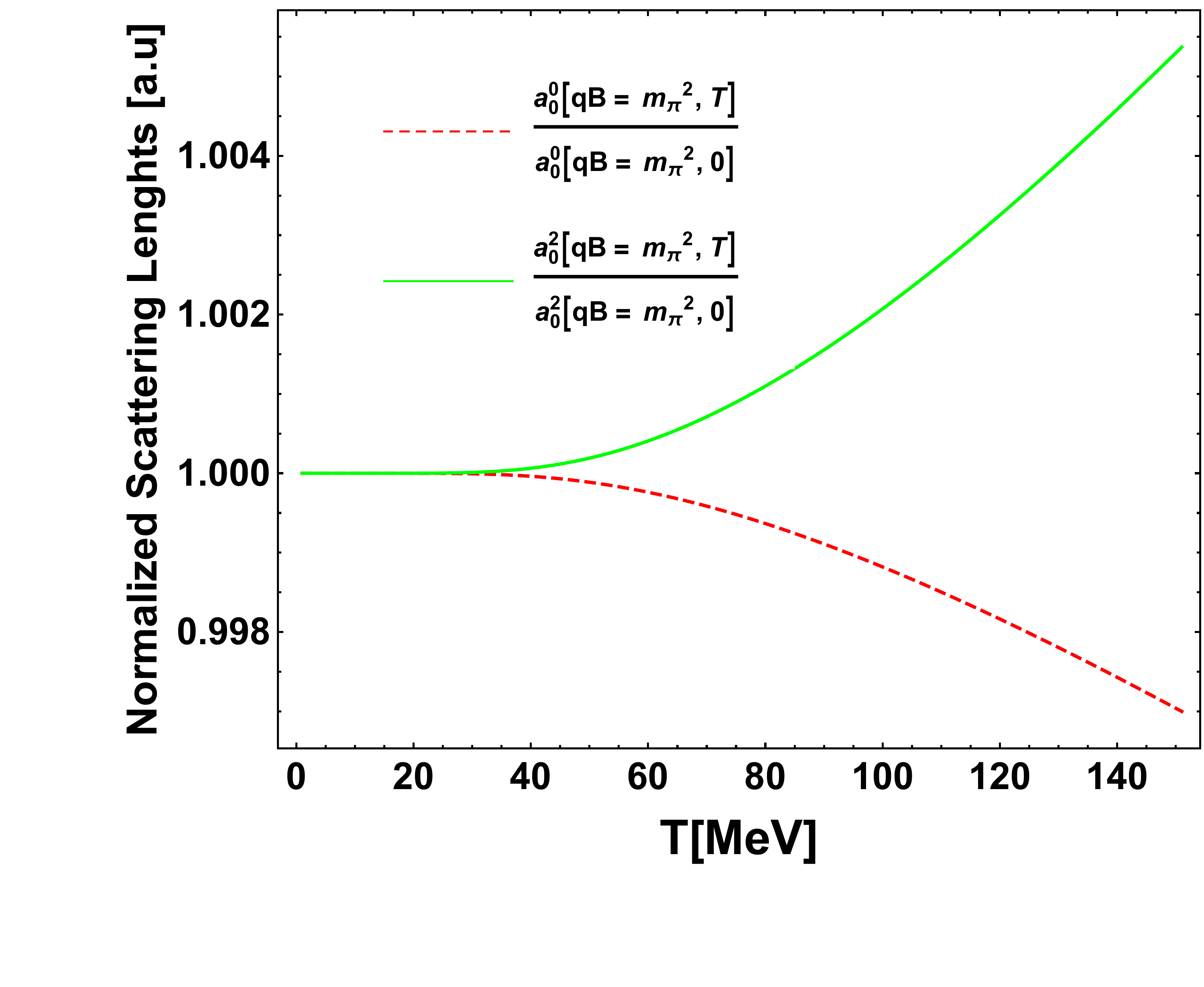}}
\caption{(Color online) The scattering parameters $a_0^{0}(B,T)/a_0^{0}(B,0)$ (dashed) and $a_0^{2}(B,T)/a_0^{2}(B,0)$ (solid) are displayed as a function of the temperature for $qB=m_{\pi}^2$.}
\label{plot2}
\end{figure}

\begin{figure}[h!]
{\centering
\includegraphics[scale=0.25]{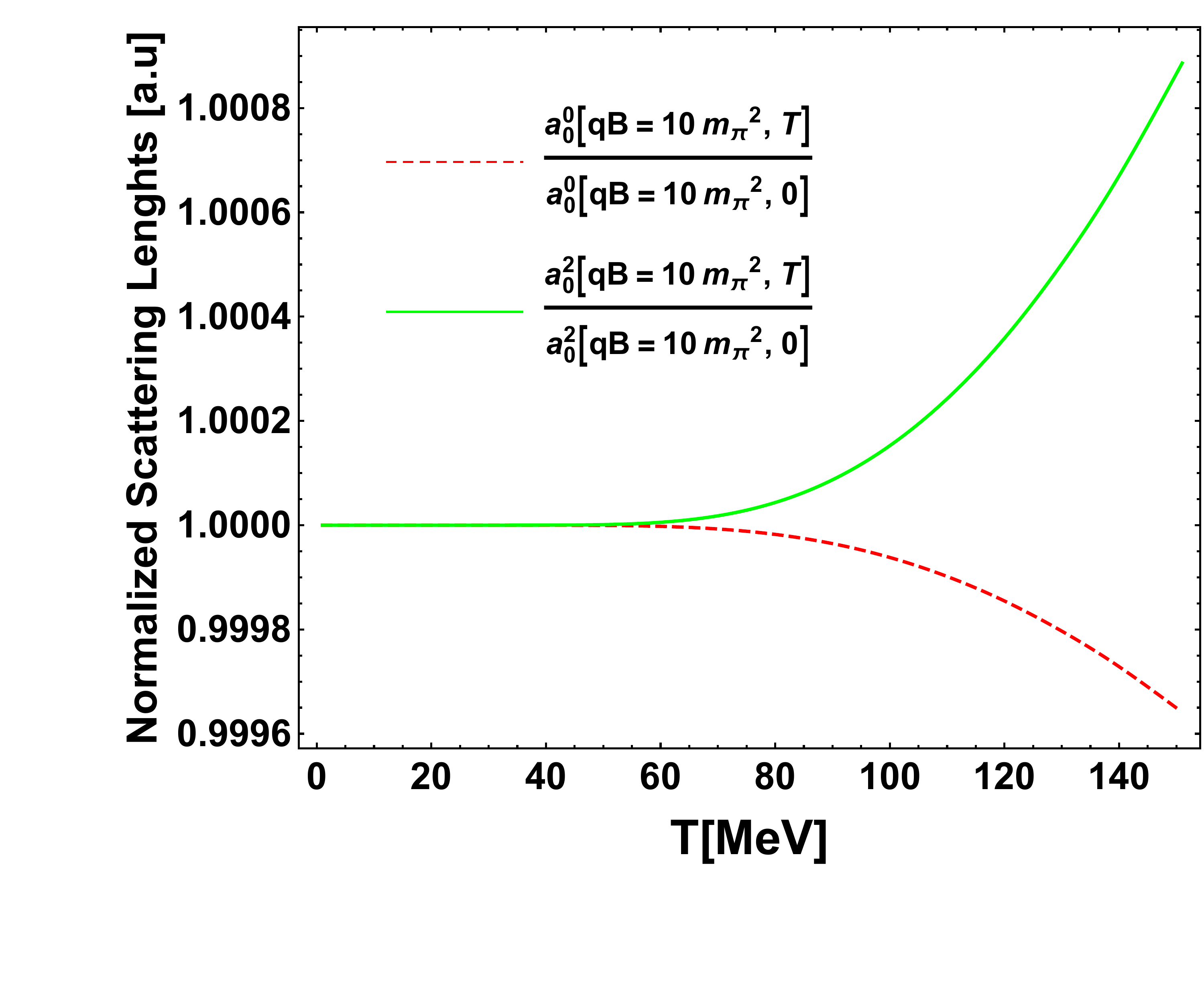}}
\caption{(Color online) The scattering parameters $a_0^{0}(B,T)/a_0^{0}(B,0)$ (dashed) and $a_0^{2}(B,T)/a_0^{2}(B,0)$ (solid) are displayed as a function of the temperature for $qB=10 m_{\pi}^2$.}
\label{plot3}
\end{figure}

\section*{ACKNOWLEDGMENTS}

 M. Loewe acknowledges support from FONDECYT (Chile) under grants No. 1170107, No. 1150471, No. 1150847 and ConicytPIA/BASAL (Chile) grant No. FB0821, E. M. acknowledges support from FONDECYT under grant No. 1190361 and R. Zamora would like to thank support from CONICYT FONDECYT Iniciaci\'on under grant No. 11160234.

\appendix
\section{Matsubara sums and integrals over $k_3$}
\begin{figure}[h!]
{\centering
\includegraphics[scale=0.58]{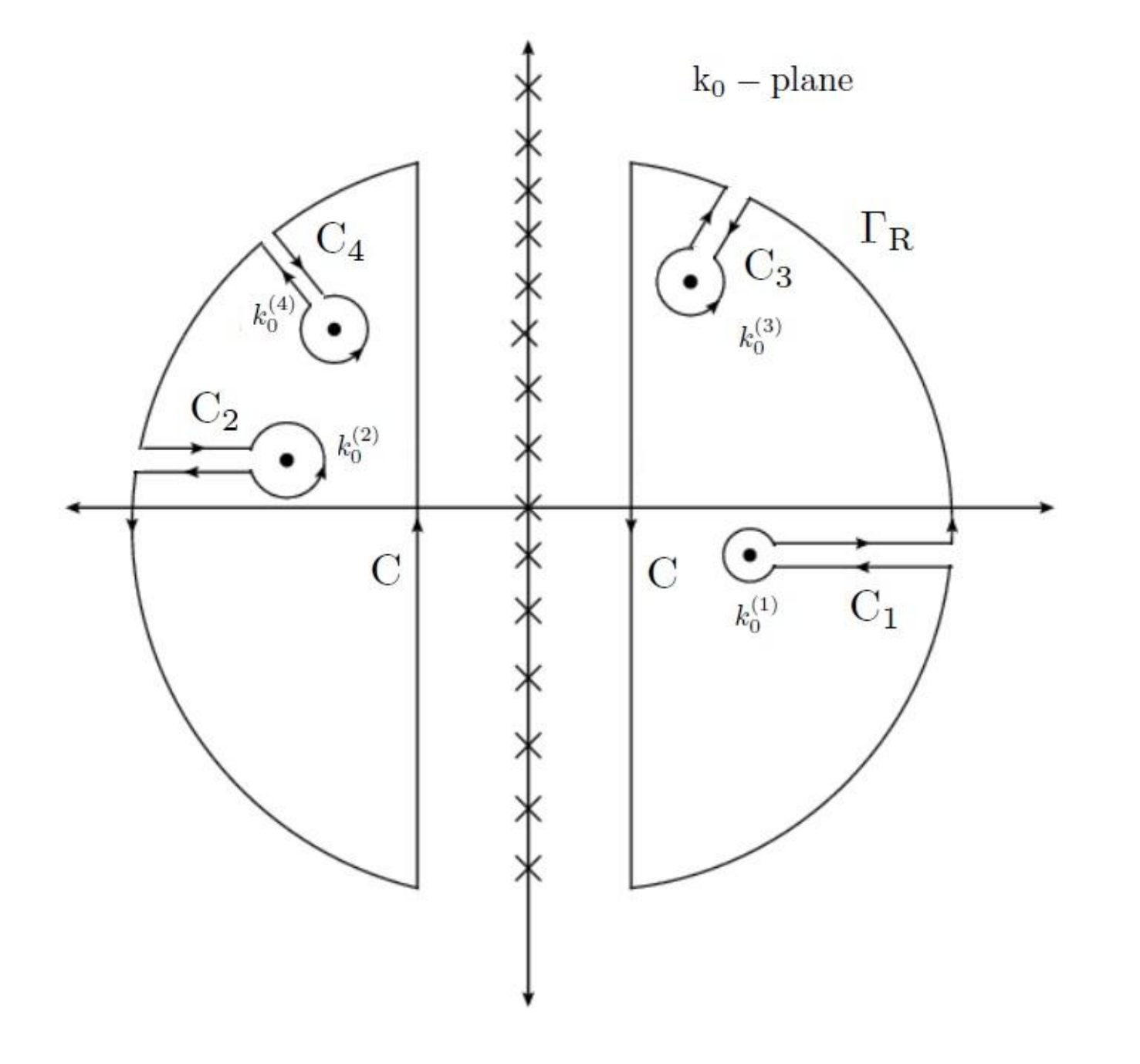}}
\caption{Integration contour in the complex $k_0$-plane}.
\label{contorno}
\end{figure}

Here we present in detail the calculation of the Matsubara sums involved in Eq.(\ref{eq_I1_reduced}) of the main text. Using the definition of the ``parallel" propagators Eq.(\ref{eq_prop_Land}),
we have
\begin{eqnarray}
G_{l,l}(T,i\nu_m) &=&\int \frac{dk_3}{2\pi}\frac{i}{\beta}\sum_{n\in Z} i\Delta_{l}^{B}(i\omega_n,k_3) \nonumber \\
&\times& \i\Delta_{l}^{B}(i\omega_n- 2 i \nu_m,k_3)\nonumber\\ 
&=& i \frac{i^2}{2\pi}\int_{-\infty}^{+\infty} dk_3 f(T,E_l(k_3),i\nu_m)
\label{eq_ap_prop}
\end{eqnarray}
where we have defined the expression
\begin{eqnarray}
f(T,E_l(k_3),i\nu_m) = \frac{1}{\beta}\sum_{n\in Z}  \frac{1}{A(i\omega_n,k_3)C(i\omega_n,k_3)}, \nonumber \\
\label{eq_f_def}
\end{eqnarray}
with 
\begin{eqnarray}
A(i\omega_n,k_3)&=&(i\omega_n)^2 - E_l(k_3)^2 + i\epsilon,\nonumber\\
C(i\omega_n,k_3)&=&(i\omega_n - 2 i \nu_m)^2- E_l(k_3)^2 + i\epsilon,
\end{eqnarray}
and $E_l(k_3) = \sqrt{k_3^2 + m_{\pi}^2 + qB(2l + 1)}$.
The Matsubara sum can be evaluated, as usual, by constructing a contour integral on the complex $k_0$-plane,
\begin{eqnarray}
\oint_{C \oplus C_1 \oplus C_2 \oplus C_3 \oplus C_4 } dk_0 \frac{n_B(k_0)}{A(k_0,k_3) C(k_0,k_3)} 
\label{a4}
\end{eqnarray}
where the contour is depicted in Fig.(\ref{contorno}). Here, the integrand contains as a factor the Bose-Einstein distribution,
\begin{eqnarray}
n_B(k_0) = \left( \exp(\beta k_0) - 1\right)^{-1}
\end{eqnarray}
that possesses infinitely-many simple poles at the Matsubara frequencies $k_0 = i\omega_n = i 2\pi n/\beta$, with
residue
\begin{eqnarray}
\lim_{k_0\rightarrow i\omega_n} \frac{k_0 - i\omega_n}{e^{i\omega_n}\exp(\beta (k_0 - i\omega_n)) - 1} = \frac{1}{\beta}.
\end{eqnarray}
In addition, the integrand in Eq.(\ref{a4}) possesses four simple
poles at $k_0^{(1,2)}=\pm E_l(k_3) \mp i\epsilon'$ and $k_0^{(3,4)}=2 i \nu_m \pm E_l(k_3) \mp i\epsilon'$, i.e., two of them located on the positive imaginary plane, while the other
two are located on the negative imaginary plane. By direct application of the residue theorem, we have
\begin{widetext}
\begin{align}
\oint dk_0 \frac{n_B(k_0)}{A(k_0,k_3) C(k_0,k_3)} = 0 = 2\pi i \left(
\sum_{j=1}^{4} n_B(k_0^{(j)})\lim_{k_0\rightarrow k_0^{(j)}}\frac{(k_0 - k_0^{(j)})}{A(k_0,k_3) C(k_0,k_3)}
+ \frac{1}{\beta}\sum_{n \in Z}  \frac{1}{A(i\omega_n,k_3) C(i\omega_n,k_3)}
\right).
\end{align}
\end{widetext}
Clearly, the second term within the parenthesis is precisely the function $f_l(T,k_3,i\nu_m)$ defined in Eq.(\ref{eq_f_def}), and hence we
have

\begin{eqnarray}
&&f(T,E_l(k_3),i\nu_m) =  -\sum_{j=1}^{4} n_B(k_0^{(j)})\nonumber \\
&\times& \lim_{k_0\rightarrow k_0^{(j)}}\frac{(k_0 - k_0^{(j)})}{A(k_0,k_3) C(k_0,k_3)}\nonumber\\
&=& \frac{\left[n_B(E_l(k_3)) - n_B(-E_l(k_3))\right]}{4i \nu_m E_l (k_3) }\nonumber\\
&\times&\left\{\frac{1}{2 E_l(k_3) - 2 i \nu_m} -  \frac{1}{2 E_l(k_3) + 2 i \nu_m}\right\}, \nonumber\\
\label{eq_f}
\end{eqnarray}
where we used the elementary property of the Bose distribution $n_B(z + i\nu_m) = n_B(z)$. Moreover, we also have
$n_B(-z) = -1 - n_B(z)$. Therefore, $n_B(E_l(k_3)) - n_B(-E_l(k_3) = 2 n_B(E_l(k_3)) + 1 = \coth(\beta E_l(k_3)/2)$, and we finally obtain
\begin{align}
f(T,E_l(k_3),i\nu_m) &=  \frac{\coth\left(\beta E_l(k_3)/2\right)}{8i \nu_m E_l (k_3) }\nonumber\\
&\times\left\{\frac{1}{ E_l(k_3) -  i \nu_m} -  \frac{1}{ E_l(k_3) +  i \nu_m}\right\}.
\label{eq_f_2}
\end{align}
Let us define the functions
\begin{eqnarray}
g(T,E,i\nu_m) = \frac{\coth\left(\beta E/2 \right)}{E[E - i\nu_m]},
\end{eqnarray}
such that
\begin{eqnarray}
f(T,E_l(k_3),i\nu_m) &=& \frac{1}{8 i\nu_m}\bigl[ g(T,E_l(k_3),i\nu_m) \nonumber \\
&-& g(T,E_l(k_3),-i\nu_m)\bigr].
\end{eqnarray}
Now we calculate the integral over $k_3$. Inserting Eq.(\ref{eq_f}) into Eq.(\ref{eq_ap_prop}), we have
\begin{eqnarray}
G_{l,l}(T,i\nu_m) &=& -\frac{i}{8 i\nu_m}\int_{-\infty}^{\infty}\frac{dk_{3}}{2\pi}\left(g(T,E_l(k_3),i\nu_m)\right.\nonumber\\
&&\left. - g(T,E_l(k_3),-i\nu_m) \right)\nonumber\\
&=& \frac{1}{4 i\nu_m} \Im m\, \int_{-\infty}^{\infty}\frac{dk_{3}}{2\pi} g(T,E_l(k_3),i\nu_m). \nonumber \\
\label{eq_ap_prop2}
\end{eqnarray}

\section{Spectral density}
As defined in the main text, here we present the analytical calculation for the spectral density $\rho(E)$. By definition,
we have
\begin{eqnarray}
\rho(E) =  \sum_{l=0}^{\infty} \int_{-\infty}^{\infty} \frac{dk_3}{2 \pi} \delta(E - E_l (k_3)).
\end{eqnarray}
The condition enforced by the argument of the delta function possesses two algebraic solutions, i.e. $k_3 = \pm \sqrt{E^2 - m_{\pi}^2 - q B (2 l +1)}$, but only
the positive one possesses support under the integration domain,
\begin{eqnarray}
\delta(E - E_l (k_3)) = \frac{E}{|k_3|} \left[
\delta\left(k_3 + \sqrt{E^2 - m_{\pi}^2 - q B (2 l +1)}\right)
\right.\nonumber\\
\left.+ \delta\left(k_3 - \sqrt{E^2 - m_{\pi}^2 - q B (2 l +1)}\right)
\right].\nonumber\\
\end{eqnarray}
By carrying out the integration explicitly, we obtain the expression
\begin{eqnarray}
\rho(E) = \frac{E}{\pi} \sum_{l=0}^{\infty} \frac{\Theta\left(E - \sqrt{m_{\pi}^{2} + qB (2l +1)} \right)}{\sqrt{E^2 - m_{\pi}^2 - q B (2l + 1)}}.
\end{eqnarray}
The infinite sum can be calculated, after noticing that the Heaviside step function restricts the upper limit of the Landau level $0 \le l \le l_{max}(E)$ for each
given value of the energy, through the condition,
\begin{eqnarray}
l_{max}(E) = \Bigg\lfloor\frac{E^2 - m_{\pi}^2 - qB}{2 q B}  \Bigg\rfloor,
\end{eqnarray}
with $\lfloor z \rfloor$ the integer part of $z$.
Therefore, from this definition, the analytical expression for the spectral density becomes
\begin{eqnarray}
\rho(E) &=& \frac{E}{\pi\sqrt{2 q B}}\Theta\left(E - \sqrt{m_{\pi}^{2} + qB } \right)\nonumber\\
&&\times\sum_{l=0}^{l_{max}(E)} \frac{1}{\sqrt{\frac{E^2 - m_{\pi}^2 - qB}{2 q B}-l}}.
\end{eqnarray}

It is now convenient to organize the finite sum differently, by defining $\bar{l} = l_{max}(E) - l$. Clearly, we have $0 \le \bar{l} \le l_{max}(E)$. Therefore, the expression above can be written in the equivalent form
\begin{eqnarray}
\rho(E) &=& \frac{E}{\pi\sqrt{2 q B}}\Theta\left(E - \sqrt{m_{\pi}^{2} + qB } \right)\nonumber\\
&&\times \sum_{\bar{l}=0}^{l_{max}(E)} \frac{1}{\sqrt{\frac{E^2 - m_{\pi}^2 - qB}{2 q B}-l_{max}(E) + \bar{l}}}.
\end{eqnarray}
\bigskip
By applying the identity that follows from the definition of the Hurwitz Zeta function, for $n \in \mathbb{N}_0$,
\begin{eqnarray}
\sum_{l = 0}^{n}\frac{1}{\left( z + l\right)^s} = \zeta(s,z) - \zeta(s,z+n+1),
\end{eqnarray}
we can reduce the density of states to the analytical expression
\begin{eqnarray}
\rho(E) &=& \frac{E}{\pi\sqrt{2 q B}}\Theta\left(E - \sqrt{m_{\pi}^{2} + qB } \right)\nonumber\\
&&\times \left\{
\zeta\left(\frac{1}{2},\frac{E^2 - m_{\pi}^2 - qB}{2 q B}-l_{max}(E) \right)\right.\nonumber\\ 
&&\left.- \zeta\left(\frac{1}{2},\frac{E^2 - m_{\pi}^2 - qB}{2 q B} + 1 \right)
\right\}.
\end{eqnarray}

\end{document}